\documentclass[aps,pra,multicolumn,showpacs,superscriptaddress,groupedaddress]{revtex4}
\usepackage{graphicx}
\usepackage[toc,page,title,titletoc,header]{appendix}

\newcommand{\Tr}{\text{Tr}}

\begin{document}
\title{Information Flow, Non-Markovianity and Geometric Phases}
\author{S. L. Wu}
\affiliation{School of Physics and Optoelectronic Technology,\\
Dalian University of Technology, Dalian 116024 China}
\author{X. L. Huang}
\affiliation{School of Physics and Optoelectronic Technology,\\
Dalian University of Technology, Dalian 116024 China}
\affiliation{School of Physics and Electronic Technology, \\
Liaoning Normal University, Dalian, 116029, China}
\author{L. C. Wang}
\affiliation{School of Physics and Optoelectronic Technology,\\
Dalian University of Technology, Dalian 116024 China}
\author{X. X. Yi}
\email{yixx@dlut.edu.cn}
\affiliation{School of Physics and Optoelectronic Technology,\\
Dalian University of Technology, Dalian 116024 China}

\begin{abstract}
Geometric phases and information flows of a two-level system coupled
to  its environment are calculated and analyzed. The information
flow is defined as a cumulant of changes in trace distance between
two quantum  states, which is similar to the measure for
non-Markovianity given by Breuer. We obtain an analytic relation
between the geometric phase and the information flow for pure
initial states, and a numerical result for mixed initial states. The
geometric phase  behaves differently depending on whether there are
information flows back to the two-level system from its environment.
\end{abstract}

%\keywords{The geometric phase, Breuer's measure, Memory effect, The
%weakly coupling limit}
 \maketitle

\section{Introduction}

Geometric phase has not been noticed for a long time until
Pancharatnam's study  \cite{Pancharatnam} and Berry's discovery
\cite{Berry}. Shortly, Simon gave a geometric interpretation of this
kind of phase in the language of differential geometry and fibre
bundles \cite{Simon1983}. Since then there was a keen interest in
holonomy effect in quantum theory, which leads to many extensions of
the geometric phase, including the geometric  phase acquired in a
non-adiabatic and cyclic evolution by Aharonov and Anandan
\cite{AAnonadibatic}, in a non-adiabatic and non-cyclic evolution by
Samel and Bhandari \cite{SBnoncyclic} and in a non-adiabatic,
non-cyclic and non-unitary evolution by Mukunda \cite{Mukunda}. All
those investigations were focused on pure state. For a practical
quantum system, however, its state would be  mixed  due to the
unavoidable coupling to its environment. This motivates the study on
the geometric phase for mixed
states\cite{Uhlmann,mixedsjoqvist,mixedtong,TongDMGP}, which was
defined by \cite{TongDMGP},
\begin{widetext}
\begin{eqnarray}
\Phi_{\text{GP}}(t)=\texttt{Arg}\left(\sum_{i=1}^N\sqrt{\epsilon_i(0)\epsilon_i(t)}\langle\psi_i(0)|
\psi_i(t)\rangle
\exp\left\{-\int_0^{t}\langle\psi_i(t')|\dot{\psi}_i(t')\rangle
dt'\right\}\right),\label{TongGP}
\end{eqnarray}
\end{widetext}
where $\epsilon_i(t)$ and $|\psi_i(t) \rangle$ are the eigenvalues
and the corresponding eigenstates of the density matrix,
respectively. Throughout this paper, we will use this definition to
study the geometric phase.

On the other hand, it is complicated  to exactly describe the
dynamics of open systems due to the huge number of variables in an
environment. In order to simplify the description, one could make
some approximations such as the weakly coupling  and the Markovian
approximation. Under these approximations, we can obtain a Markovian
master equation that describes the dynamics of the open system
without memory of its history. However, many systems exhibit strong
non-Markovian effect and can not be described by the Markovian
master equation. It is then interesting   to study the geometric
phase in a non-Markovian dynamics, and establish the relation
between the geometric phase and the non-Markovianity of the
dynamics.

The non-Markovianity may be defined in many ways
\cite{breuer,use1,use2,extend}, for instance, in Ref.\cite{breuer}
the authors  proposed  a scheme to quantify the degree of the
non-Markovianity   based on the trace distance of two quantum
states\cite{breuer}, and in Ref.\cite{use1} the non-Markovianity was
measured by exploiting the specific traits of quantum correlations.
The effects of non-Markovianity on geometric phase has been
considered by several works, e.g. Refs.\cite{Huang} and \cite{chen}.
In this paper, we will establish a relation between the geometric
phase and the information flows. We divide the information flow into
two types, i.e., the information flow from the open system into the
environment $\mathcal M$ (call forward information flow) and the
information flow back from the environment to the open system
$\mathcal N$ (call backward information flow). The backward
information flow $\mathcal N$ in fact is a modified measure of
non-Markovianity given by Breuer and his co-workers. This paper is
organized as follows. In Sec.II, we present  a definition for the
information flow based on the trace distance, then we establish the
relation between the information flow and the geometric phase for
pure initial states. The case of mixed initial states is considered
in Sec.III, where the geometric phase and the information flow are
calculated numerically. Finally, we present conclusion and
discussions in Sec.IV.

\section{Information flow and  geometric phase for pure initial states}

\subsection{A measure for information flow}

Here we first recall the measure for non-Markovianity defined by
Breuer\cite{breuer}. This definition is based on the so-called trace
distance between two states $\rho_1$ and $\rho_2$
\begin{eqnarray}
D(\rho_1,\rho_2)=\frac{1}{2} \Tr||\rho_1-\rho_2||,\label{distance}
\end{eqnarray}
where $||A||=\sqrt{AA^\dag}$. For a two-level system, this trace
distance is equal to one half of  the ordinary Euclidean distance
between the two states on the Bloch sphere, i.e.
$D(\rho_1,\rho_2)=\frac{1}{2} |\vec r_1-\vec r_2|$, where $\vec r_j$
is the Bloch vector for state $\rho_j$. The   change rate of the
trace distance can be represented as
\begin{eqnarray}
\sigma(\rho_1(t),\rho_2(t))=\frac{d}{dt}D(\rho_1(t),\rho_2(t)).\label{ddistance}
\end{eqnarray}%
When $\sigma<0$, $\rho_1(t)$ and $\rho_2(t)$ approach  to each other
in the dynamics  and this can be understood as information flow from
the system to the environment; when $\sigma>0$, $\rho_1(t)$ and
$\rho_2(t)$ is away from each other in the time evolution, and this
can be interpreted as information flow  back to the system, which is
treated as  a typical character of the non-Markovianity. As shown in
Refs.\cite{breuer} and \cite{breuer2}, one can define a measure of
non-Markovianity $\mathcal N_B(t)$ for a quantum process $\Psi(t)$
by  maximizing  over all initial states $(\rho_1(0),\rho_2(0))$ of
total gain  of the trace distance, namely,
\begin{eqnarray}
\mathcal N_B (\Psi)=\max\limits_{\rho_{1,2}(0)} \int_{\sigma>0}
\sigma(\rho_1(t),\rho_2(t))\,dt .\label{measuren1}
\end{eqnarray}%
The measure of non-Markovianity only characterizes the gain of the
trace distance in the dynamics. To describe the loss and gain of the
trance distance uniformly, we extend the concept of information flow
in the following. In fact, one of the information flows defined
below is a modified  measure of non-Markovianity in
Eq.(\ref{measuren1}).  We choose a steady state as $\rho_2$ in Eq.
(\ref{distance}) and call it standard state. This choice makes the
maximization easy, however, it can not measure all non-Markovian
dynamics. Fortunately, this simplified measure of non-Markovianity
is in agreement with the measure given in \cite{breuer} in our case.

Now we  define  $\mathcal M$ and $\mathcal N$ to measure the
information  gain and lose for a given initial state $\rho_1(0)$ in
the dynamics $\Psi(t)$. We will refer this information gain and loss
uniformly as  formation flows.
\begin{eqnarray}
\mathcal N (\Psi)=\int_{\sigma>0} \sigma(\rho_1(t),\rho_2)\,dt
,\label{measuren}
\end{eqnarray}
and
\begin{eqnarray}
\mathcal M (\Psi)=-\int_{\sigma<0} \sigma(\rho_1(t),\rho_2)\,dt
.\label{measurem}
\end{eqnarray}%
Obviously, the above two measures satisfy,
\begin{eqnarray}
D(\rho_1(t),\rho_2)=D(\rho_1(0),\rho_2)&+\mathcal N
(\rho_1(t),\rho_2) -\mathcal M (\rho_1(t),\rho_2)&.\label{rmeasuren}
\end{eqnarray}%
The difference between $\mathcal N(t)$ and $\mathcal N_B$ is as
follows. $\mathcal N_B$ is the maximum information flow back to the
system in the dynamics. Hence it does not depend on the initial
condition. However $\mathcal{N}(t)$ is the information flow back to
system with respect to the standard state $\rho_2(0)$ in the
interval $(0,~t)$  (assumed that a steady state for quantum process
$\Psi(t)$ exists, and the standard state $\rho_2(0)$ is exactly the
steady state as before). This simplification is true in the
situation considered in this paper, i.e., a two-level system
coupling to vacuum electromagnetic fields at zero temperature with
standard state $\rho_2$ in the Bloch sphere representation as
$\vec{r}_2=(0,0,-1)$.

With these definition and notations, we will discuss the
relationship between the geometric phase and the information flows
between the open system and its environment.

\subsection{The geometric phase for pure initial states}

Consider a two-level system coupled to its environment at zero
temperature. The general form of the density matrix can be expressed
as $\rho(t) = \frac{1}{2} (1+\vec{r}(t) \cdot \vec{\sigma})$, where
$\vec\sigma=(\sigma_x,\sigma_y,\sigma_z)$ is the Pauli matrices, and
$\vec r(t)=r(t)\cdot(\sin\theta(t)\cos\phi(t),
\sin\theta(t)\sin\phi(t), \cos\theta(t))$ is the  Bloch vector.  For
pure initial states $|\vec r(0)|^2=1$, while for mixed initial
states $ |\vec r(0)|^2<1 $. It is easy to obtain the instantaneous
eigenvalues of the above density matrix $\rho(t)$ as,
\begin{eqnarray}
\epsilon_\pm(t) = \frac{1}{2} ( 1 \pm |\vec r(t)| ).\label{epsilon}
\end{eqnarray}%
It is obvious that for the pure initial states, the eigenvalue
$\epsilon_-(t=0)=0$, which means that  the eigenstate corresponding
to the eigenvalue $\epsilon _-(t)$ gives no contribution to the
geometric phase. This simplifies our calculation and helps us to
obtain an analytic result for the geometric phase. The eigenstates
corresponding to the eigenvalues in Eq.(\ref{epsilon}) can be
written as,
\begin{eqnarray}
|{\psi_+(t)}\rangle{=}\left(\begin{array}{cc}
            \sin\frac{\theta(t)}2\\
            \cos\frac{\theta(t)}{2}e^{i \phi(t)}
          \end{array}\label{eigenstate}
          \right) ,
|{\psi_-(t)}\rangle{=}\left(\begin{array}{cc}
            -\cos\frac{\theta(t)}2\\
            \sin\frac{\theta(t)}{2}e^{i \phi(t)}
          \end{array}\label{eigenstate}
          \right) .
\end{eqnarray}%
Now substitute Eq.(\ref{epsilon}) and Eq.(\ref{eigenstate}) into the
Eq.(\ref{TongGP}) with an assumption that  $\phi=\omega_0 t+ \phi_0$
(where $\omega_0$ and $\phi_0$ are constants, this is reasonable for
different kinds of master equation \cite{Huang,chen}), the geometric
phase with  pure initial states (their Bloch vector is $\vec{r}(0)$)
can be obtained as
\begin{eqnarray}
 \Phi_{\text{GP}}=-\int_0^T
\omega_0\cos^2\frac{\theta(t)}{2}\,dt ,\label{ggeometrcialphase}
\end{eqnarray}%
where we set $T=\frac{2n\pi}{\omega_0}$ $(n=1,2,3,...)$, i.e., a
multiple of the quasi-period $\frac{2\pi}{\omega_0}$.

Next, we  establish the relationship between the geometric phase and
the information flows $\mathcal M(t)$ and $\mathcal N(t)$. Keeping
the relationship Eq.(\ref{rmeasuren}) between the $\mathcal N(t)$
and $\mathcal M(t)$ in mind, we obtain the geometric phase acquired
by the system,
\begin{widetext}
\begin{eqnarray}
\Phi_{\text{GP}}{=}{-}\int_0^T
\omega_0[\frac{1}2{+}\frac{r_z(t)}{2\sqrt{4
[D(0){+}\mathcal{N}(t){-}\mathcal{M}(t)]^2{-}2r_z(t){-}1}}]\,dt
.\label{dgeometricalphase}
\end{eqnarray}%
\end{widetext}
where $r_z(t)=r(t)\cos\theta(t)$ is the z-component of $\vec r(t)$
and $D(0)=D(\rho_1(0),\rho_2)$. It is shown that when the system is
closed, i.e. $\cos\theta(t)=\cos\theta_0$ and $r(t)=1$, where
$\theta_0$ is initial polar angle on the Bloch sphere, and setting
$T=2\pi/\omega_0$, Eq.(\ref{dgeometricalphase}) reduced to the
well-known form, $\Phi_{\text{GP}}^{(closed)}=-\pi(1+\cos\theta_0)$,
that is the geometric phase acquired by a two-level quantum system
in a rotating magnetic field. For a Markovian process, $\mathcal
N(t)$ is always zero and $\mathcal M(t)$  increases  with time until
it approaches $D(0)$. In this case, the geometric phase is only
influenced by the information flow to  the environment. When we
consider the non-Markovian effects, the situation is  more
complicated, and the information  flow  $\mathcal N(t)$  back to the
open system has a link to the geometric phase given by
Eq.(\ref{dgeometricalphase}). We will discuss it numerically in
Sec.{\rm III}.

\section{The geometric phase for mixed state}
In this section,  we will study the geometric phase of a two-level
system with mixed initial states. Because it is difficult to get an
analytical result like Eq.(\ref{dgeometricalphase}) for the
information flows and the geometric phase, we here numerically and
perturbatively establish a relation between the geometric phase and
the information flow. The perturbation is carried out to the first
order in the coupling constant, while the numerical results are for
a wide range of coupling constants. Two types of master equation,
the time-local master equation and the memory kernel master equation
with exponential memory, will be  considered.

\subsection{The time-local master equation}\label{sec3a}

Let us consider a two-level system interacting with a vacuum field
at zero temperature whose spectral density is  Lorentzian
\cite{breuer3,breuer4,projection1},
\begin{eqnarray}
J(\omega)=\frac{1}{\pi}\frac{W^2\lambda}{(\omega_0-\omega)^2+\lambda^2}.
\end{eqnarray}%
Here $W$ is the coupling constant  between the system and the
environment, $\omega_0$ is the atomic transition frequency which is
of the time scale $\tau_0\sim\omega_0^{-1}$, and $\lambda$ is the
spectral width of the coupling that is connected to the environment
correlation time,  $\tau_B\sim\lambda^{-1}$. The dynamics of this
system is governed by the following master equation (namely the
time-local master equation) \cite{projection1},
\begin{eqnarray}
\dot\rho(t)=-i\Delta(t)[\sigma_+\sigma_-,\rho(t)]+
\Gamma(t)[2\sigma_-\rho(t)\sigma_+-\sigma_+\sigma_-\rho(t)-\rho(t)\sigma_+\sigma_-],\label{timelocal}
\end{eqnarray}%
where $\sigma_\pm$ are the pauli  operators and the parameters
\begin{eqnarray}
\begin{array}{cc}
  \Delta(t)=-\Im[\frac{\dot c(t)}{c(t)}], & \Gamma(t)=-\Re[\frac{\dot
  c(t)}{c(t)}],
\end{array}
\end{eqnarray}%
play the role of  Lamb shift and decay rate for the system,
respectively.  Here $c(t)$ can be calculated by means of Laplace
transform as,
\begin{eqnarray}
c(t)=\exp\left[\frac{-(\lambda+i\omega_0)t}{2}\right]
\left(\cosh\frac{\Omega
t}{2}+\frac{\lambda}{\Omega}\sinh\frac{\Omega
t}{2}\right),\label{ct}
\end{eqnarray}%
with $\Omega=\sqrt{\lambda^2-4 W^2}$. We note that the rate
$R=\frac{W}{\lambda}$ indicates the strength of the
non-Markovianity. With $R<\frac 1 2$, the dynamics is called
time-dependent Markovian,  while for $R>\frac 1 2$ the dynamics is
non-Markovian. Here we assume again that the initial state of the
open system  is
\begin{eqnarray}
\rho(0)=\frac12(1+\vec r_0\cdot \vec{\sigma}), \label{xxx}
\end{eqnarray}
where $\vec r_0=r_0\cdot(\sin\theta_0 \cos\phi_0, \sin\theta_0
\sin\phi_0, \cos\theta_0 )$ and $|\vec r_0|^2<1$.
 With  this initial condition, the  density matrix of
the system at time $t$  can be obtained from the master equation
Eq.(\ref{timelocal})
\begin{eqnarray}
\rho(t)=\frac{1}{2}\left(\begin{array}{cc}
     (1+r_0\cos\theta_0)|c(t)|^2 & r_0\sin\theta_0\exp(i\phi_0)c(t) \\
     r_{0}\sin\theta_0\exp(-i\phi_0)c^*(t) & 2-(1+r_0\cos\theta_0)|c(t)|^2
   \end{array}
   \right),\label{tldensity}
\end{eqnarray}%
The eigenvalues and the eigenstates of the reduced density matrix
Eq.(\ref{tldensity}) can be easily obtained  as
\begin{eqnarray}
\epsilon_\pm(t) &=& \frac{1}{2} ( 1 \pm r(t) )\label{tleigenvalue},\nonumber\\
|\psi_+(t)\rangle&=&\left(\begin{array}{cc}
            \sin\frac{\theta_t}2 \\
            \cos\frac{\theta_t}2e^{i(\omega_0t+ \phi_0)}
          \end{array}\label{tleigenstate}
          \right),\\
|\psi_-(t)\rangle&=&\left(\begin{array}{cc}
            -\cos\frac{\theta_t}2 \\
            \sin\frac{\theta_t}2e^{i(\omega_0t+ \phi_0)}
          \end{array}\label{tleigenstate}
          \right),\nonumber
\end{eqnarray}%
where
$\tan\theta_t=\frac{r_0\sin\theta_0|c(t)|}{(1+r_0\cos\theta_0)|c(t)|^2-1}$
and
$r(t)=\sqrt{[(1+r_0\cos\theta_0)|c(t)|^2-1]^2+r_0^2\sin^2\theta_0|c(t)|^2}$.

We can expand the geometric phase with respect to  the coupling
strength $W^2$ up to the first order (i.e., in the weak coupling
limit), that is
\begin{eqnarray}
\Phi_{\text{GP}}(T)\doteq\Phi_{\text{GP}}^{(0)}-W^2[\tan
\Phi_{\text{GP}}^{(0)}C_1(r_0,\theta_0)\kappa_1(\lambda,T)+\omega_0
\cos^{-2} \Phi_{\text{GP}}^{(0)}
C_2(r_0,\theta_0)\kappa_2(\lambda,T)],\label{tlexpend}
\end{eqnarray}%
where $\kappa_1(\lambda,T)=\frac{\partial\mid c(T) \mid^2}{\partial
W^2}\mid_{W^2=0}=\frac{1-\exp(-\lambda
T)}{\lambda^2}-\frac{T}{\lambda}$,
$\kappa_2(\lambda,T)=\int_0^T\frac{\partial|c(t)|^2}{\partial
W^2}|_{W^2=0}\,d
t=\frac{T}{\lambda^2}+\frac{1}{\lambda^3}[\exp(-\lambda
T)-1]-\frac{T^2}{2\lambda},$ and $\Phi_{\text{GP}}^{(0)}$ is the
geometric phase acquired under the unitary evolution with mixed
initial states,  $\Phi_{\text{GP}}^{(0)}=\arctan(r_0\tan(-i\pi
(1+\cos \theta_0)))$. The parameter $C_i(r_0,\theta_0)(i=1,2)$ is a
constant relative to the initial condition given by
\begin{eqnarray}
&C_1(r_0,\theta_0)=\frac{1}{4}(r_0+r_0\cos^2\theta_0+2\cos\theta_0),\nonumber\\
&C_2(r_0,\theta_0)=\frac{1}{r_0}(1+\frac{r_0\sin^2\theta_0\cos\theta_0}{2}-\cos^2\theta_0).
\end{eqnarray}

It is interesting to calculate the trace distance between $\rho(t)$
defined in Eq.(\ref{tldensity}) and the standard state in the weak
coupling limit. Substituting Eq.(\ref{tldensity}) into
Eq.(\ref{distance}), and expanding the trace distance up to the
first order in  $W^2$, we have
\begin{eqnarray}
D(t)\doteq D(t)|_{W^2=0}+\frac{W^2}{4D(t)|_{W^2=0}}
\times[r_0^2(1+\cos^2\theta_0)+2r_0\cos\theta_0]\frac{\partial\mid
c(t) \mid^2}{\partial W^2}\mid_{W^2=0}, \label{dexpend}
\end{eqnarray}%
where $D(t)\mid _{W^2=0}=\sqrt{r_0^2+1+2r_0\cos\theta_0}/2$ and
$\frac{\partial\mid c(t) \mid^2}{\partial
W^2}\mid_{W^2=0}=\frac{1-\exp(-\lambda
t)}{\lambda^2}-\frac{t}{\lambda}$. $D(t)|_{W^2=0}$ is the trace
distance at time $t$ under unitary evolution, it is not difficult to
prove that $D(t)\mid _{W^2=0}=D(0)$. Then according to
Eq.(\ref{rmeasuren}), we have
\begin{eqnarray}
\mathcal N(t)-\mathcal M(t)\doteq\frac{W^2}{4D(t)|_{W^2=0}}
\times[r_0^2(1+\cos^2\theta_0)+2r_0\cos\theta_0] \frac{\partial\mid
c(t) \mid^2}{\partial W^2}\mid_{W^2=0}. \label{weakdistance}
\end{eqnarray}%
If coupling strength is very weak, there is no information flow back
into the system in a quasi-period, i.e. $\mathcal N(T)=0$. So it is
straightforward to obtain,
\begin{eqnarray}
\mathcal M(t)\doteq-\frac{W^2}{4D(t)|_{W^2=0}}
\times[r_0^2(1+\cos^2\theta_0)+2r_0\cos\theta_0]\frac{\partial\mid
c(t) \mid^2}{\partial W^2}\mid_{W^2=0}. \label{mweakdistance}
\end{eqnarray}%
This result tells us that $\mathcal M(t)$  increases  monotonically
with the increasing of $1/\lambda$. Comparing this result with
Eq.(\ref{tlexpend}), one may find that the dependence  of $\mathcal
M(T)$ and $\Phi_{\text{GP}}(T)$ on the spectral width $\lambda$ is
almost the same  for  mixed initial states in the weak coupling
limit.

\begin{figure}
\includegraphics[scale=0.5]{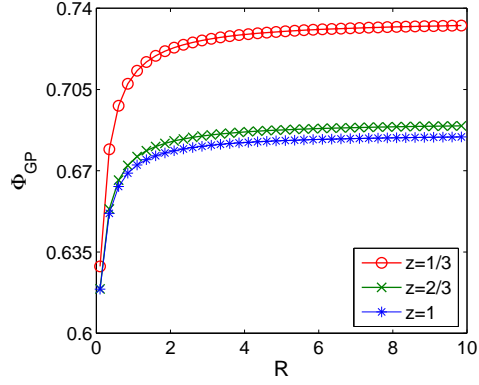}
\caption{(Color online) The geometric phase (in units of $\pi$) as a
function of the parameter $R$ with different $z$. The coupling
strength is $W=0.1\omega_0$ and $\omega_0=1$. The azimuthal angle of
the two components of the initial state are chosen as
$\vartheta_0=\pi/4$, $\varphi_0=\frac{\pi}{3}$. }\label{w01phase1}
\end{figure}

\begin{figure}
\includegraphics[scale=0.5]{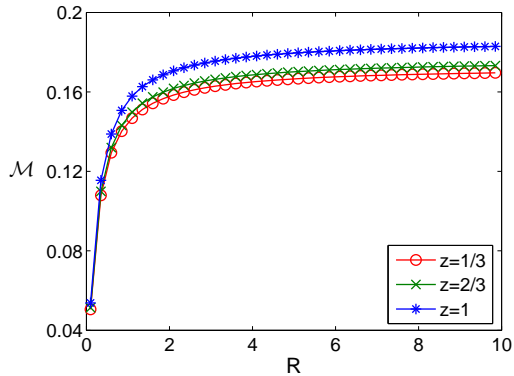}
\caption{(Color online) The information flow $\mathcal M$ as a
function of the parameter $R$ with different $z$. The parameters are
the same as in Fig.\ref{w01phase1} }\label{w01measure1}
\end{figure}

Numerical results for the geometric phase and the information flow
$\mathcal M$ under the weak coupling limit is shown in
Fig.\ref{w01phase1} and Fig.\ref{w01measure1}. The initial states
are chosen as ,
\begin{eqnarray}
\rho(0)=\frac{1-z}{2}\mathbf{I}+z\left|\xi\rangle\langle\xi\right|,\label{initialcondition}
\end{eqnarray}%
where
$|\xi\rangle=\cos\vartheta_0|0\rangle+\sin\vartheta_0\exp(i\varphi_0)|1\rangle$
is a pure state, $z\in[0,1]$ and $\mathbf{I}$ is a $4\times4$
unitary matrix. For $z=0$, the density matrix is the maximumlly
mixed state, while they reduce to a pure one in the case of $z=1$.
In the language of  Bloch vector, the initial state
Eq.(\ref{initialcondition}) can be represented  as
\begin{eqnarray}
r_0=z, \theta_0=2\vartheta_0,  \phi_0=\varphi_0.
\end{eqnarray}


\begin{figure}
\includegraphics[scale=0.5]{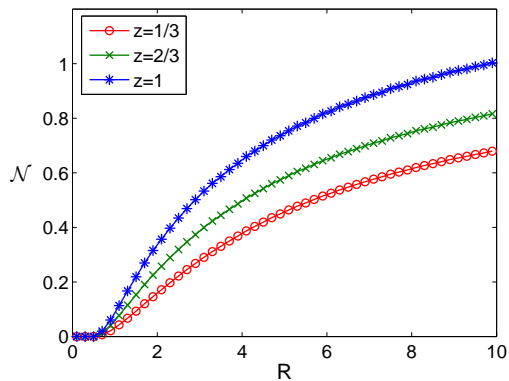}
\caption{(Color online) The information flow $\mathcal N$ as a
function of the parameter $R$ with different $z$. Here the coupling
strength $W=\omega_0$ and $\omega_0=1$. The initial states of the
open system are chosen as Fig.\ref{w01phase1}. }\label{w1measure1}
\end{figure}

\begin{figure}
\includegraphics[scale=0.5]{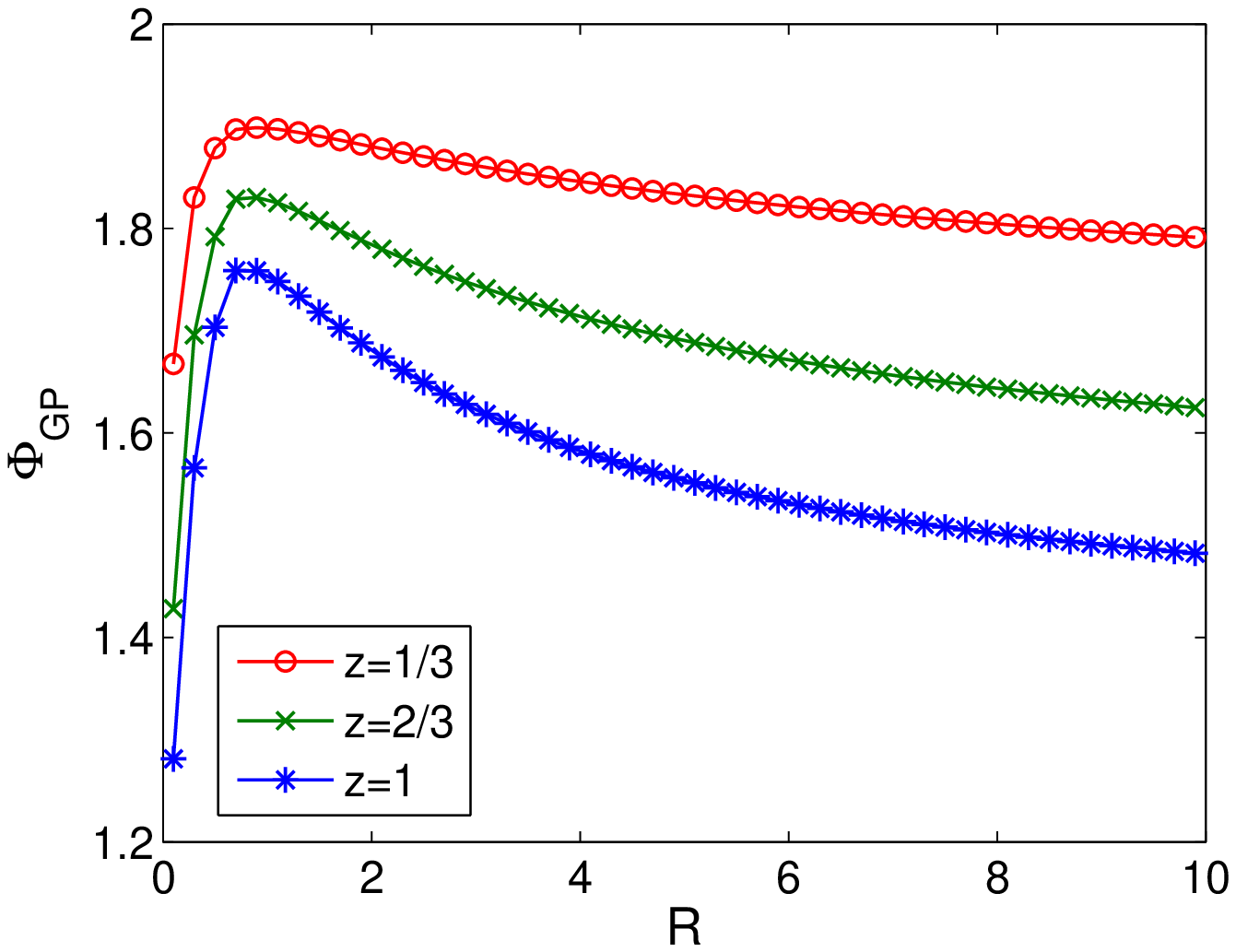}
\caption{(Color online) The geometric phase as a function of the
parameter $R$ with different $z$. Here the coupling strength
$W=\omega_0$ and $\omega_0=1$. The initial state of the open system
are chosen as Fig.\ref{w01phase1}. }\label{w1phase1}
\end{figure}

\begin{figure}
\includegraphics[scale=0.53]{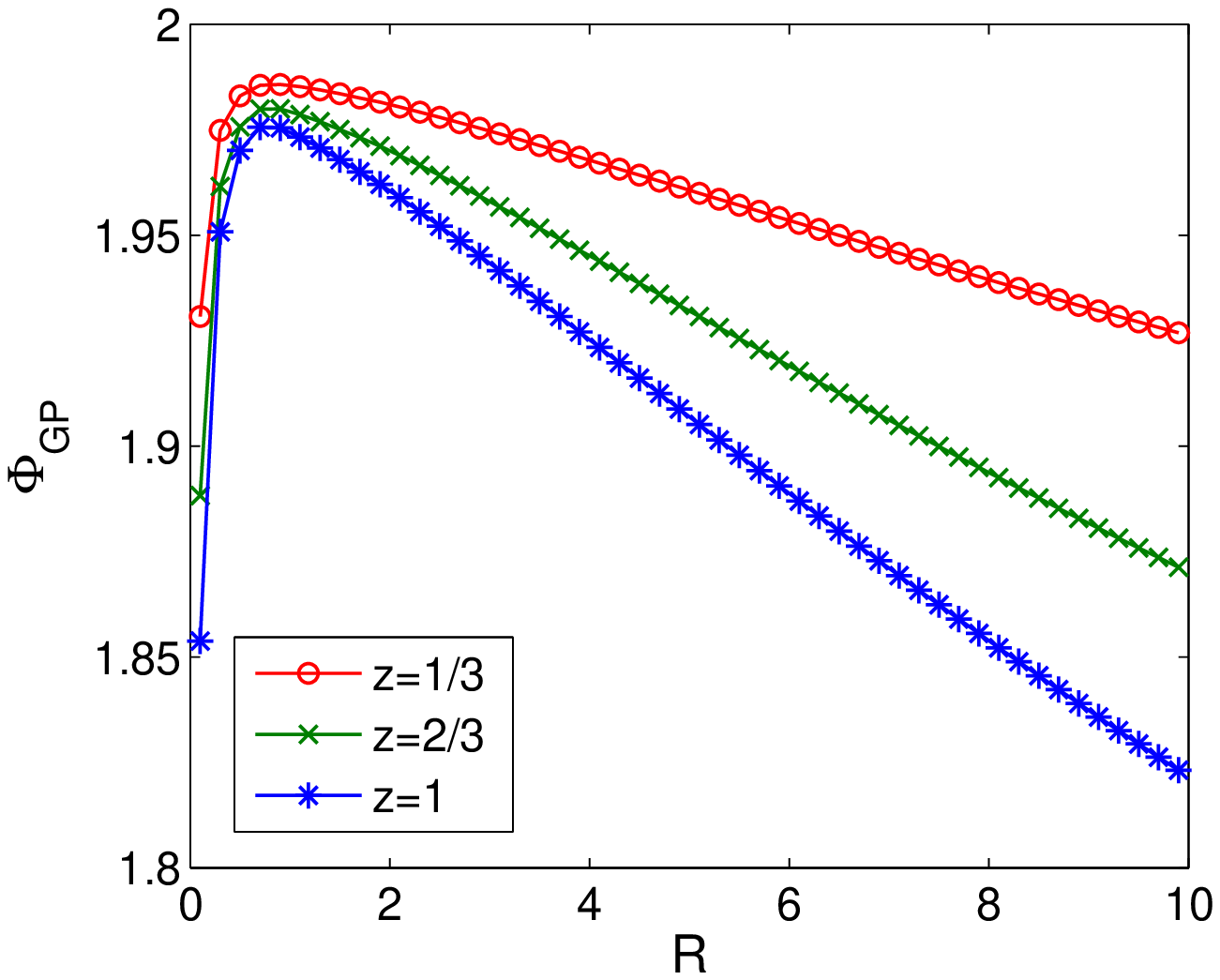}
\caption{(Color online) The geometric phase as a function of  $R$
with different $z$. Here the coupling strength $W=10\omega_0$ and
$\omega_0=1$. The initial states of the open system are the same as
in Fig.\ref{w01phase1}. }\label{w10phase1}
\includegraphics[scale=0.5]{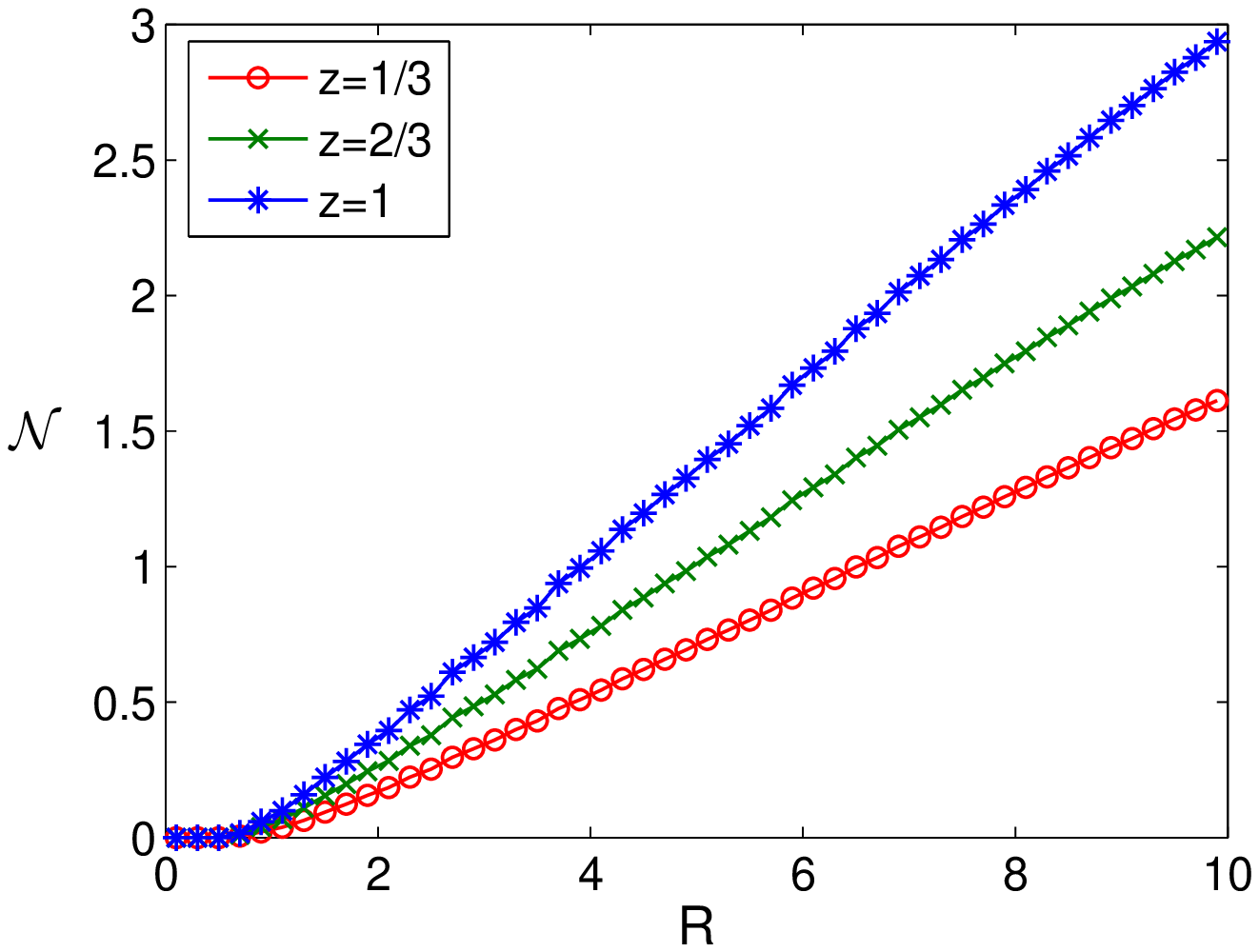}
\caption{(Color online) The information flow $\mathcal N$ as a
function of  $R$ with different $z$. Here the coupling strength
$W=10\omega_0$ and $\omega_0=1$. The initial states of the open
system are the same  as in Fig.\ref{w01phase1}. }\label{w10measure1}
\end{figure}

 Now we establish the relation between the geometric phase and the
information flow with  different coupling strengths  one by one.
E.g., $W=0.1\omega_0$, $W=\omega_0$ and $W=10\omega_0$ will be
chosen to explore the geometric phase and the information flow,
numerical results are shown in
Fig.(\ref{w01phase1})--(\ref{w10measure1}). Fig.\ref{w01phase1} and
Fig.\ref{w01measure1} show the geometric phase $\Phi_{\text{GP}}$
and the information flow $\mathcal M$ as a function of $R$  with
different parameter $z$ for  weak coupling ($W=0.1 \omega_0$). The
geometric phase $\Phi_{\text{GP}}$ is plotted in units of $\pi$. We
can see from Fig.\ref{w01phase1} that the geometric phase increases
monotonically with the parameter $R$. In this case, the coupling
strength $W$ is small enough so that there is no information flowing
back into the system in a quasi-period. For  pure initial states,
i.e. $z=1$ , comparing Fig.\ref{w01phase1} with
Fig.\ref{w01measure1}, we find that the larger the geometric phase
is, the more information flow to the environment, this is confirmed
by Eq.(\ref{dgeometricalphase}). Because $\mathcal N(R)$ is always
zero in a quasi-period, the geometric phase mainly depends on
  $\mathcal M(R)$.

With increase  of the coupling strength $W$, the correlation time of
the environment $\tau_B$ approaches  to the time $T$, which
indicates that $\mathcal N(R)>0$ in a quasi-period.  In
Fig.\ref{w1measure1} and Fig.\ref{w1phase1}, we plot the information
flow $\mathcal N(R)$ and the geometric phase $\Phi_{\text{GP}}$ as a
function of $R$ for $W=\omega_0$. Here we only focus on  the
information flow $\mathcal N$, because it characterizes the
non-Markovianity of the open system and describe the backward
information flow. 

Comparing Fig.\ref{w1measure1} with Fig.\ref{w1phase1}, we may find
that, when the information flows  back to the system, the geometric
phase decreases with the increase of $\mathcal N(R)$ and, in the
region of $\mathcal N(R)=0$, the behavior of the geometric phase is
similar with the case in the  weak coupling limit. This indicates
the backward information flow  (i.e., information flow back to the
open system) affects the geometric phase acquired by the open
system. This phenomena can be understood as follows: For this
time-local master equation, when the information flows back to the
system, the Bloch vector moves toward to the north pole of the
sphere, then the geometrical phase which is interpreted as the solid
angle in Bloch sphere decreases.

When $W$ is very large, the system will drop into steady state after
time $T$.  In this case, by comparing the geometric phase
$\Phi_{\text{GP}}$ and the information flow $\mathcal N$, we can see
that, when $\mathcal N(R)=0$ (as shown in Fig.\ref{w10measure1}),
the behavior of the geometric phase (as shown in
Fig.\ref{w10phase1}) is very similar to the case of $W=0.1\omega_0$
and $W=\omega_0$ with $\mathcal N(R)=0$. With the increase of $R$,
the information flow back into the open system and the geometric
phase decreases with $R$, this is very similar to the case when
$W=\omega_0$ in the region of $\mathcal N(R)>0$. Based on  these
 observations, we conclude that for a dynamics described by  the
time local master equation, if the geometrical phase is inversely
proportional  to $R$, the dynamics must be non-Markovian. In other
words, the non-Markovianity can be reflected in  the geometrical
phase to a certain extent. This conclusion are both valid for pure
and mixed initial states.

It is seemingly  that the point (in the $R$ axis)  where the
geometric phase arrives its extremum is exactly  the point where
$\mathcal N(T)$ begin to increase (in the following we will call it
as the critical  point), however, by carefully examination, we find
that this is not the case.  According to
Eq.(\ref{dgeometricalphase}) and the definition of trace distance,
we can clarify  that the critical  point is the very point where the
integrand of the geometric phase reaches its minimum (a  detail of
proof can be found in  the Appendix). Moreover, the integrand of the
geometrical phase behaves similarly  with the information flow
$\mathcal N(t)$.
\begin{figure}
\includegraphics[scale=0.53]{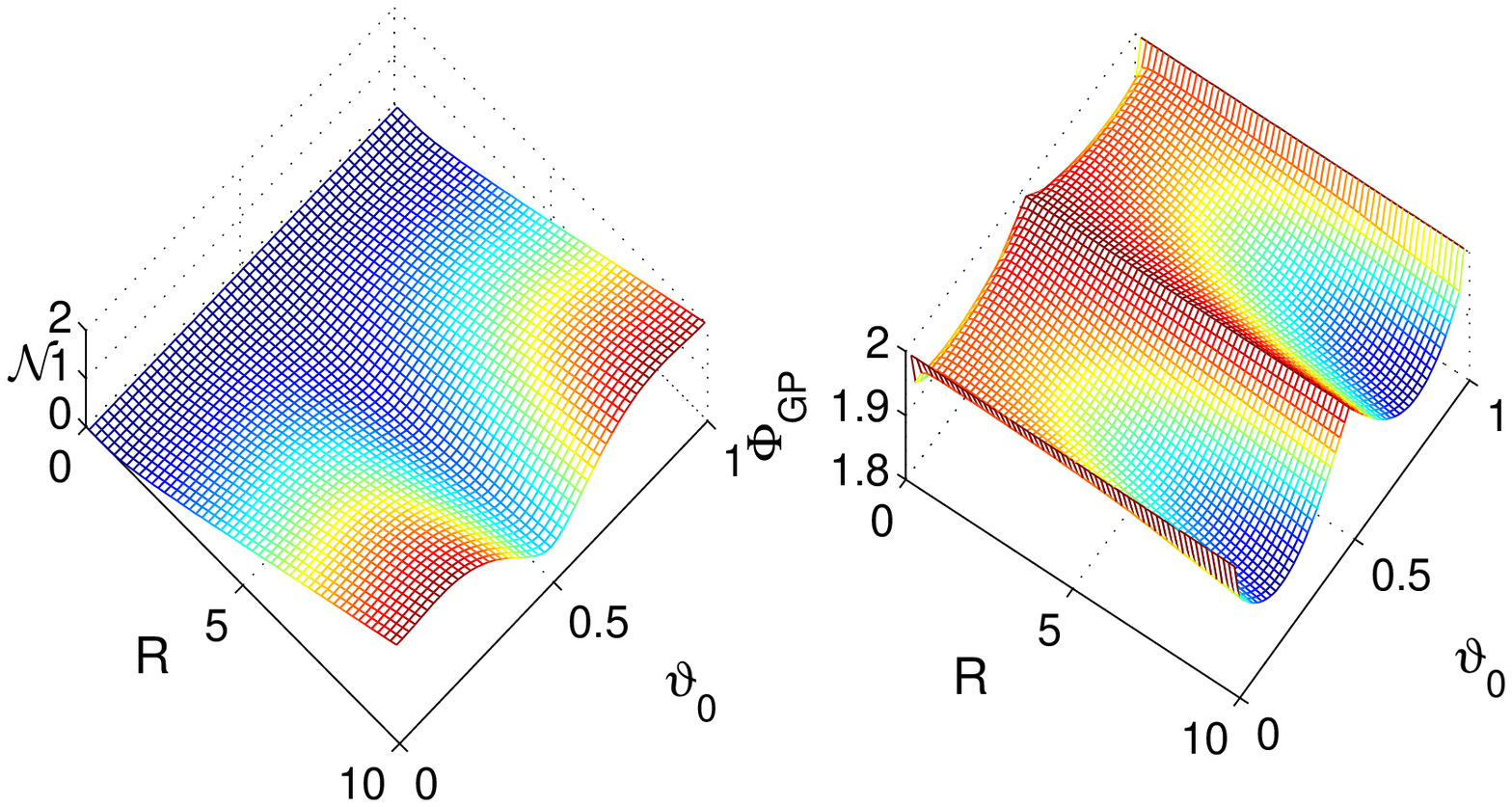}
\caption{(Color online) (a)The information flow $\mathcal N$ and (b)
the geometric phase $\Phi_{\text{GP}}$ as a function of $R$ and the
initial polar angle $\vartheta_0$ (in units of $\pi$). Here the
coupling strength is $W=10\omega_0$ and $\omega_0=1$. The initial
states of the open system are chosen as $z=1/2$ and
$\varphi_0=\pi/6$. }\label{tc}
\end{figure}

The information flow $\mathcal N(T)$ and the geometric phase
$\Phi_{\text{GP}}$ as a function of $\vartheta_0$ and $R$ are
plotted in Fig.\ref{tc}.  When $\mathcal N(T)>0$, it can be seen
from the figure that the more the information flows back, the
smaller the geometric phase; in the region of $\mathcal N(T)=0$, no
matter how to choose the initial azimuthal  angle $\vartheta_0$, the
geometric phase increase with $R$, which is exactly the finding of
our analytical analysis.  When $\vartheta_0=\pi/2$, both $\mathcal
N(T)$ and $\Phi_{\text{GP}}$ reach its extremum, in this case, the
geometric phase do not change with $R$ and $\Phi_{\text{GP}}=2\pi$,
but the information flow changes indeed. This can be understood as
follow: when $\vartheta_0=\pi/2$, $\sin2\vartheta_0=0$,  the initial
density matrix is diagonal, i.e.
$\rho(0)=\text{diag}\{(1-z)/2,(1+z)/2\}$, with this diagonal density
matrix, the geometric phase is always equal to $2\pi$ and $\mathcal
N(T)$ changes with $R$ since the information exchange between the
system and the environment varies with $R$. The situation remains
unchanged for the cases where $\vartheta_0=0$ and $\vartheta_0=\pi$.
Furthermore we find that, although $\Phi_{\text{GP}}=2\pi$ for both
$\vartheta_0=\pi/2$ and $\vartheta_0=\pi$, the information flows
$\mathcal N(T)$ are completely different. This can be explained as
the difference in the initial states, which are
$\rho(0)=\text{diag}\{(1-z)/2,(1+z)/2\}$ for $\vartheta_0=\pi/2$ and
$\rho(0)=\text{diag}\{(1+z)/2,(1-z)/2\}$ for $\vartheta_0=\pi$. By
the definition of the geometric phase, it depends on the spectrum of
the density matrix, which are the same for the initial states,
leading to the same geometric phase acquired in the dynamics. But
the information flow $\mathcal N(T)$ for $\vartheta_0=\pi$ is larger
than that for $\vartheta_0=\pi/2$, this is because the information
flow was defined as the distance between the actual state and the
standard state of the open system, which are different for the
initial states.

\subsection{The memory kernel master equation with exponential memory}

Now we consider the geometric phase of an open two-level system
governed by the memory kernel master equation with exponential
memory. Here, we just apply this model to calculate the geometric
phase but do not discuss the positivity of the master equation in
detail.

In the interaction picture, the integro-differential master equation
with memory kernel can be expressed as
\begin{eqnarray}
\dot\rho(t)=\int_0^tK(t')\mathcal{L}\rho(t-t')\,d
t',\label{nzmastereq}
\end{eqnarray}%
where $\mathcal{L}$ is Liouvillian superoperator which takes the
form
\begin{eqnarray}
\mathcal{L}\rho=\frac{1}{2}\gamma_0(2\sigma_-\rho\sigma_+
-\sigma_+\sigma_-\rho-\rho\sigma_+\sigma_-),
\end{eqnarray}
and $\gamma_0$ is the dissipation rate, $K(t)$ represents the memory
effect  called  Shabani-Lidar memory kernel\cite{Barnett,Maniscalco}
\begin{eqnarray}
K(t)=\gamma \exp(-\gamma t).
\end{eqnarray}
We call $\tau_R=\frac{1}{\gamma}$ the memory time. It is not
difficult to solve this integro-differential equation by the Laplace
transform with the initial condition Eq.(\ref{xxx}). In the
Schr\"{o}dinger picture, the solution is
\begin{widetext}
\begin{eqnarray}
\rho(t)&=&\frac{1}{2}\left(\begin{array}{cc}
     (1+r_0\cos\theta_0)\xi(C,t) & r_0\sin\theta_0e^{-i(\omega_0t+\phi_0)}\xi(\frac{C}{2},t) \\
     r_0\sin\theta_0e^{i(\omega_0t+\phi_0)}\xi(\frac{C}{2},t) & 2-(1+r_0\cos\theta_0)\xi(C,t)
   \end{array}
   \right) \label{nzdensity},
\end{eqnarray}%
\end{widetext} where
\begin{eqnarray}
&&\xi(C,\tau)=e^\frac{-\tau}{2} [\cosh(\frac{\Omega
\tau}{2})+\frac{1}{\Omega}\sinh(\frac{\Omega \tau}{2})]\label{naxi},\nonumber \\
&&\Omega=\sqrt{1-4C}\nonumber.
\end{eqnarray}%
The parameters $C$ and $\tau$ are defined as
$C=\frac{\gamma_0}{\gamma}$ and $\tau=\gamma t$.

\begin{figure}
\includegraphics[scale=0.48]{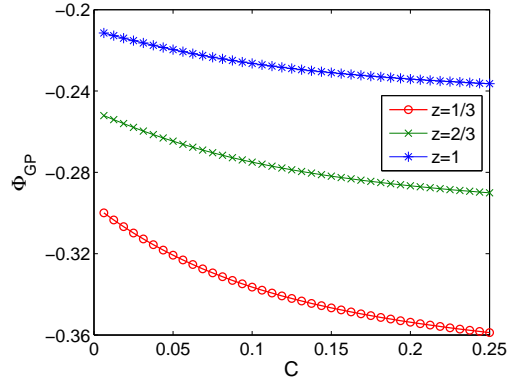}
\caption{(Color online) The geometric phase for the memory kernel
master equation with exponential memory kernel as a function of the
parameter $C$ with  different $z$. Here the dissipation constant
takes $\gamma_0=0.1\omega_0$ and $\omega_0=1$. We choose the same
initial states as in Fig.\ref{w01phase1} for the open system to plot
this figure. }\label{nzp}
\end{figure}

\begin{figure}
\includegraphics[scale=0.52]{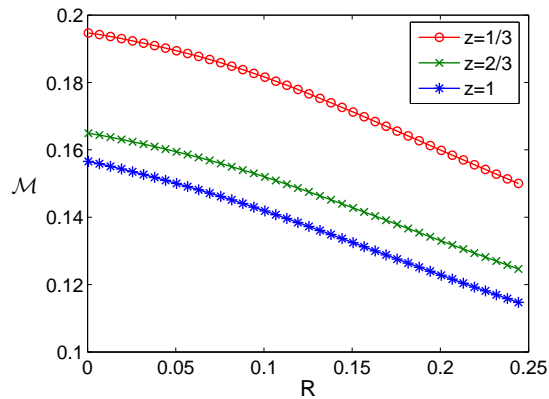}
\caption{(Color online) The information flow $\mathcal M$ for memory
kernel master equation with exponential memory kernel as a function
of  $C$ with different $z$, where the dissipation constant is
$\gamma_0=0.1\omega_0$ and $\omega_0=1$. The initial states of the
open system are the same as in Fig.\ref{w01phase1}. }\label{nzm}
\end{figure}

By the same procedure as in Sec.\ref{sec3a}, we establish relation
between the geometric phase and the information flow, which is found
to be similar to that in the last section. Hence, the conclusions
for the relation between the geometric phase and the information
flow hold true for open systems described by memory kernel master
equation.  See the numerical results  shown in Fig.\ref{nzp} and
Fig.\ref{nzm}.


For the memory kernel master equation with exponential memory
kernel, it is well known that, in some region of  parameter $C$, the
master equation Eq.(\ref{nzmastereq}) may lead to non-positive
 density matrix. In Ref.\cite{Maniscalco}, the positivity for
a density matrix has been discussed: When $C>\frac{1}{4}$, the the
memory kernel master equation with exponential memory is not valid,
because the second perturbation used to drive the master equation
does not suit this case. Recently, Breuer et al. has checked that
this kind of master equation does not own memory effect when the
positivity of the density matrix is conserved  \cite{Mazzola}. When
$C<\frac{1}{4}$, there is no information flow back into the open
system, no matter how to choose the parameter $\gamma_0$, so we
choose $\gamma_0=0.1\omega_0$ and plot the geometric phase and the
information flow $\mathcal M(C)$ in this region. Comparing
Fig.\ref{nzp} with Fig.\ref{nzm}, we can find the link  between the
information flow $\mathcal M(C)$ and the geometric phase is the same
as we found in Sec.\ref{sec3a}.

\section{conclusion and discussions}
In summary, we have discussed the information flows and the
geometric phase in different non-Markovian process. For an open
two-level system with pure initial states, an analytic relation
between the information flow and the geometric pahse has been given
in terms of Bloch vector. For mixed initial states, two kinds of
master equation, namely time-local master equation and memory kernel
master equation with exponential memory, have been numerically
studied.  We  find that, in both cases, the information flows
influence the geometric phase directly, a relation between the
geometric phase and information flow is numerically established. An
understanding for the observation is provided.

 The forward  and backward information flow are by definition
different, but they complementarily describe the information
exchange between the environment and the system. The backward
information flow can be used to describe the Non-Markovianity of the
open system, while the forward information flow was connected with
the coherence loss. Neither forward information flow nor backward
information flow can be measured directly, indicating that the
measure of non-Markovianity defined in this way can not be directly
observed in experiment. However, mixed state geometric phases are
measurable and the feature caused by it has been
observed\cite{du03}. This motivates the establishment of the
connection between the geometric phase and the information flow.
Indeed, The finding of this paper suggests that the geometric phase
can reflect the non-Markovianity and then can serve as   a measure
of non-Markovianity for open systems.

This work is supported by NSF of China under Grant Nos. 10775023,
10935010 and 10905007.

\section*{APPENDIX}

\begin{figure}
\includegraphics[scale=0.45]{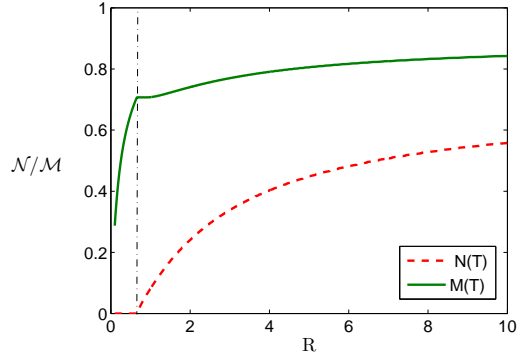}
\caption{(Color online) The measure $\mathcal N$ and $\mathcal M$ as
a function of  $R$ with a pure initial state($\vartheta_0=\pi/3$).
The coupling strength $W$ is chosen as $0.6$. This figure shows
that, at the start point (the point of crossover with the dot-dashed
line) of the measure $\mathcal N(T)$, the derivative of the measure
$\mathcal M(T)$ \label{NM} with respect to $R$ is zero.}\label{X}
\end{figure}

\begin{figure}
\includegraphics[scale=0.5]{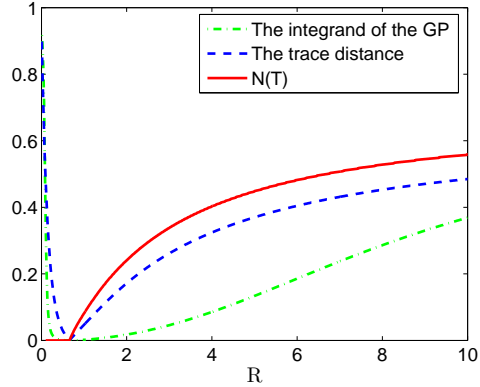}
\caption{(Color online) The integrand of the geometric phase
$A(T,R)$, the trace distance $D(T,R)$ and the measure $N(T,R)$ are
plotted in this figure. The initial state is chosen as
$\vartheta_0=\pi/3$ and the coupling strength $W$ is equal to
0.6.}\label{Y}
\end{figure}

In this appendix, we show in  detail that the critical point is
exactly the  point where $\mathcal N(t)$ begin to increase. By
Eq.(\ref{dgeometricalphase}), we write the integrand as
\begin{widetext}
\begin{eqnarray}
A(t,R)=\omega_0[\frac{1}{2}+\frac{r_z(t,R)}{2\sqrt{4D^2(t,R)-2r_z(t,R)-1}}],
\end{eqnarray}
 where $r_z(t)=r(t)\cos\theta(t)$.
To find the critical point, we take a derivative with respect to $R$
\begin{eqnarray}
\frac{\partial}{\partial R}A(t,R)=\frac{\omega_0
r_0^2\sin^2\theta_0}{4r(t,R)^3}[(1+r_0\cos\theta_0)|c(t,R)|^2+1]\frac{\partial}{\partial
R}|c(t,R)|^2,
\end{eqnarray}
noting that $r(t,R)\in[0,1]$ and
$(1+r_0\cos\theta_0)|c(t,R)|^2+1>1$, we find, when
$\frac{\partial}{\partial R}|c(t,R)|^2=0$, $A(t,R)$ must reach its
extremum. Moreover, we check the first derivative of the trace
distance with respect to $R$,
\begin{eqnarray}
\frac{\partial}{\partial
R}D(t,R)=\frac{1}{4D(t,R)}[(1+r_0\cos\theta_0)^2|c(t,R)|^2+r_0^2\sin^2\theta_0]\frac{\partial}{\partial
R}|c(t,R)|^2. \label{dderivative}
\end{eqnarray}
Obviously, since $D(t,R)\in[0,1]$ and
$(1+r_0\cos\theta_0)^2|c(t,R)|^2+r_0^2\sin^2\theta_0\geq0$, we can
see that  the trace distance arrive at its extremum if and only if
$\frac{\partial}{\partial R}|c(t,R)|^2=0$, this is exactly  the
condition for the integrand to reach its maximum. Substituting
Eq.(\ref{rmeasuren}) into Eq.(\ref{dderivative}), we obtain
\begin{eqnarray}
\frac{\partial}{\partial R} \mathcal N(t,R)-\frac{\partial}{\partial
R} \mathcal M(t,R)=\frac{1}{4D(t,R)}
[(1+r_0\cos\theta_0)^2|c(t,R)|^2+r_0^2\sin^2\theta_0]\frac{\partial}{\partial
R}|c(t,R)|^2 .
\end{eqnarray}
\end{widetext}
At the critical  point, there is no backward information flow, i.e.
$\mathcal N(t)=0$. Thus if the critical  point is the very point
satisfied $\frac{\partial}{\partial R}|c(t,R)|^2=0$, the derivative
of $\mathcal M(t)$ must be zero at this point. Because the property
of the information flows, it is difficult to obtain an analytic
results. The numerical result of $\mathcal N(T)$ and $\mathcal M(T)$
have been shown in Fig.\ref{X},  which validate our hypothesis. This
is further confirmed by Fig. \ref{Y}.

\end{document}